\documentclass{article}

\usepackage{arxiv}

\usepackage[utf8]{inputenc} 
\usepackage[T1]{fontenc}    
\usepackage{hyperref}       
\usepackage{url}            
\usepackage{booktabs}       
\usepackage{amsfonts}       
\usepackage{nicefrac}       
\usepackage{microtype}      
\usepackage{lipsum}
\usepackage{graphicx}
\usepackage{float}
\usepackage{amsmath}
\usepackage{multirow}
\graphicspath{ {./images/} }

\title{HC-Mamba: Vision MAMBA with Hybrid Convolutional Techniques for Medical Image Segmentation}

\author{
 Jiashu Xu \\
  School of Science\\
  Harbin Institute of Technology, Shenzhen\\
  \texttt{jiashu.xu04@gmail.com} \\
}

\begin{document}
\maketitle
\begin{abstract}
Automatic medical image segmentation technology has the potential to expedite pathological diagnoses, thereby enhancing the efficiency of patient care. However, medical images often have complex textures and structures, and the models often face the problem of reduced image resolution and information loss due to downsampling. To address this issue, we propose HC-Mamba, a new medical image segmentation model based on the modern state space model Mamba. Specifically, we introduce the technique of dilated convolution in the HC-Mamba model to capture a more extensive range of contextual information without increasing the computational cost by extending the perceptual field of the convolution kernel. In addition, the HC-Mamba model employs depthwise separable convolutions, significantly reducing the number of parameters and the computational power of the model. By combining dilated convolution and depthwise separable convolutions, HC-Mamba is able to process large-scale medical image data at a much lower computational cost while maintaining a high level of performance. We conduct comprehensive experiments on segmentation tasks including organ segmentation and skin lesion, and conduct extensive experiments on Synapse, ISIC17 and ISIC18 to demonstrate the potential of the HC-Mamba model in medical image segmentation. The experimental results show that HC-Mamba exhibits competitive performance on all these datasets, thereby proving its effectiveness and usefulness in medical image segmentation.
\end{abstract}


\section{Introduction}
Modern medical research is inextricably linked to the utilization of various medical images\cite{litjens2017survey}. Medical images are designed to provide an accurate visual representation of the structure and function of various tissues and organs within the human body. They assist medical professionals and scientific researchers in exploring the normal and abnormal conditions of patients in great detail, thereby serving clinical and research purposes. In both laboratory-based cutting-edge medical research and in the clinical setting, medical image analysis plays a pivotal role in facilitating scientific inference and diagnosis. \cite{chen2022recent}Automatic medical image segmentation technology has the potential to expedite pathological diagnoses, thereby enhancing the efficiency of patient care.

In recent years, a considerable amount of research on the computer-aided system for healthcare applications has been conducted\cite{muhammad2020eeg,doi2007computer,wang2023chatcad}.CNN-based and Transformer-based models have demonstrated excellent performance in a variety of vision tasks, especially in medical image segmentation. UNet\cite{ronneberger2015u}, as a representative of CNN-based models, is known for its simple structure and scalability, and many subsequent improvements are based on this U-shaped architecture. TransUnet\cite{chen2021transunet} is a pioneer in the field of Transformer-based models, it initially employs the Vision Transformer (ViT)\cite{dosovitskiy2020image} for feature extraction during the encoding phase and a Convolutional Neural Network (CNN) during the decoding phase. It demonstrates a robust capacity to capture global information. TransFuse\cite{zhang2021transfuse} integrates the parallel architectures of ViT and CNN to simultaneously capture both local and global features. Furthermore, Swin-UNet\cite{cao2022swin} integrates Swin Transformer\cite{liu2021swin} with a U-shaped architecture, representing the inaugural instance of a U-shaped model that is exclusively based on Transformer.

However, although existing models have achieved some success in feature extraction, they still face the problem of reduced image resolution and information loss due to downsampling when dealing with medical images with complex textures and structures. To address this issue, Yu F. and Koltun V.\cite{yu2015multi} proposed the technique of dilated convolution. Dilated convolution allows the model to capture a wider range of contextual information without increasing the computational cost by extending the receptive field of the convolution kernel. Because it has the ability to enhance the perception of different scale structures of images without losing image details, it is especially suitable for medical images. However, since the dilated convolution increases the perceptual field by inserting "0" between the elements of the convolution kernel, the captured features may not be coherent or accurate in some cases.

In recent times, studies based on state space models (SSMs) have attracted considerable interest from researchers \cite{goel2022s,gu2023mamba,fu2022hungry}.Building on the findings of classical SSM research\cite{kalman1960new}, modern SSMs (e.g., Mamba\cite{gu2023mamba}) not only establish long-range dependencies but also exhibit linear complexity with respect to input size. In particular, U-Mamba\cite{ma2024u} demonstrates its potential by combining SSM with CNN for the first time in the context of medical image segmentation tasks. Inspired by this, we propose HC Mamba, a model based on SSM, which integrates a variety of convolution methods optimized for medical images, in order to further demonstrate its potential in the task of medical image segmentation.

We introduce the technique of dilated convolution in the HC-Mamba model. By feeding the features generated by the dilated convolution into the SSM, the state transition capability of the SSM can be utilized to enhance the spatial correlation between the features, thus compensating for the discontinuities introduced due to the voids.

In addition, the HC-Mamba model employs depthwise separable convolutions\cite{chollet2017xception}, a convolution method that decomposes the traditional convolution operation into two parts: depthwise convolution and pointwise convolution, which significantly reduces the number of parameters and the computational power of the model. By combining dilated convolutions and depthwise separable convolutions, HC-Mamba is able to process large-scale medical image data at a much lower computational cost while maintaining a high level of performance, which is particularly important for real-time medical image processing and large-scale medical data analysis.

We conducted comprehensive experiments on segmentation tasks including organ segmentation and skin lesion , and conduct extensive experiments on Synapse, ISIC17 and ISIC18\cite{codella2019skin} to demonstrate the potential of the HC-Mamba model in medical image segmentation. The experimental results show that HC-Mamba exhibits competitive performance on all these datasets, thereby proving its effectiveness and usefulness in medical image segmentation.

In conclusion, our contribution to the field can be summarized as follows:
\begin{itemize}
    \setlength{\topsep}{2pt}
    \setlength{\parsep}{0pt}
    \setlength{\parskip}{1pt}
    \item We propose a hybrid convolution Mamba model (HC Mamba) for medical image segmentation, which combines a variety of convolution methods optimized for medical images to improve the receptive field of the model and reduce the parameters of the model.
    \item We propose the HC-SSM module for enhancing the model's ability to extract features
    \item We conducted extensive performance evaluations of the proposed model. The results show that our model has high accuracy (94.84\%), mIoU (78.42\%) and validity of DSC (87.89\%).
\end{itemize}

\section{Methods}
\subsection{Preliminaries}
\begin{figure}[H]
    \centering
    \includegraphics[width=0.75\linewidth]{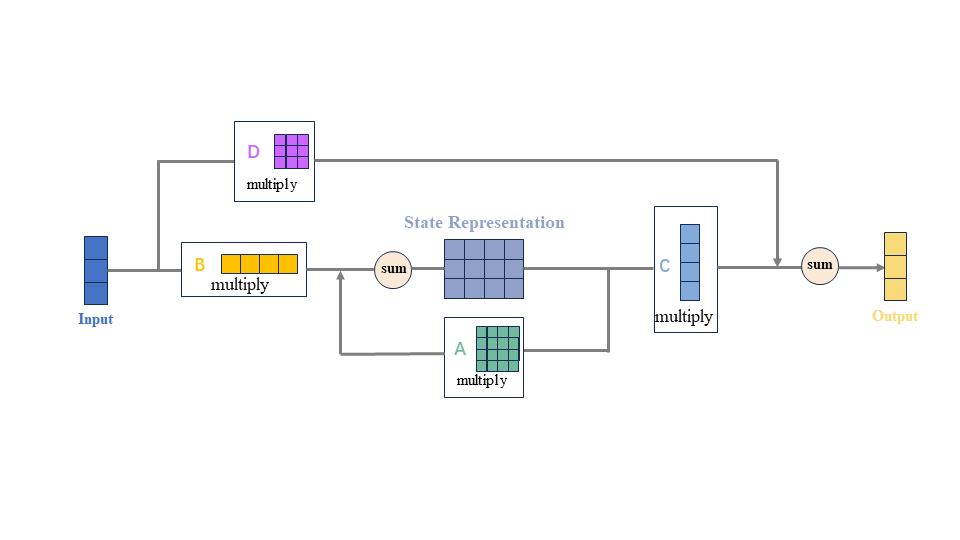}
    \caption{SSM(state space model)process diagram}
    \label{fig:1}
\end{figure}

Modern models based on State Space Models (SSM), particularly the Structured State Space Sequence Model (S4) and Mamba model, are classical continuous systems. The system maps a one-dimensional input function or sequence \( x(t) \in \mathbb{R} \) to an output \( y(t) \in \mathbb{R} \) via an implicit latent state \( h(t) \in \mathbb{R}^N \), as shown in Equation \ref{SSM}. 

\begin{equation}
\left\{ 
  \begin{aligned}
       h'(t) &= Ah(t) + Bx(t) \\ \label{SSM}
     y(t) &= Ch(t)
  \end{aligned}
\right .
\end{equation}
where, \( A \in \mathbb{R}^{N \times N} \) is the state matrix, while \( B \in \mathbb{R}^{N \times 1} \) and \( C \in \mathbb{R}^{N \times 1} \) represent the projection parameters.The process is shown in the Figure \ref{fig:1}.In the figure, the symbol D represents a skip connection, which can be understood as a transformed residual connection. Consequently, the portion of the graph that excludes D is typically designated as SSM.
To adapt these continuous systems for deep learning applications, S4 and Mamba discretize the system. Specifically, a time scale parameter, or step size \(\Delta\), is introduced, and fixed discretization rules such as Zero Order Hold (ZOH) are used to transform \( A \) and \( B \) into discrete parameters \(\hat{A}\) and \(\bar{B}\):
\begin{equation}
\left\{ 
\begin{aligned}
\hat{A} &= \exp(\Delta A) \\
\bar{B} &= \Delta A^{-1}(\exp(\Delta A) - I)\Delta B
\end{aligned}
\right .
\end{equation}
After discretization, the state space model computation can be implemented either through linear recursion:
\begin{equation}
\left\{ 
\begin{aligned}
h'(t) &= Ah(t) + Bx(t) \\
y(t) &= Ch(t)
\end{aligned}
\right .
\end{equation}
or global convolution:
\begin{equation}
\left\{
\begin{aligned}
K &= (CB, CAB, \ldots, CA^{L-1}B) \\
y &= x \ast K
\end{aligned}
\right .
\end{equation}
where, \( K \in \mathbb{R}^L \) represents a structured convolution kernel, and \( L \) denotes the length of the input sequence \( x \).

\subsection{Model structure}
\begin{figure}[H]
    \centering
    \includegraphics[width=1\linewidth]{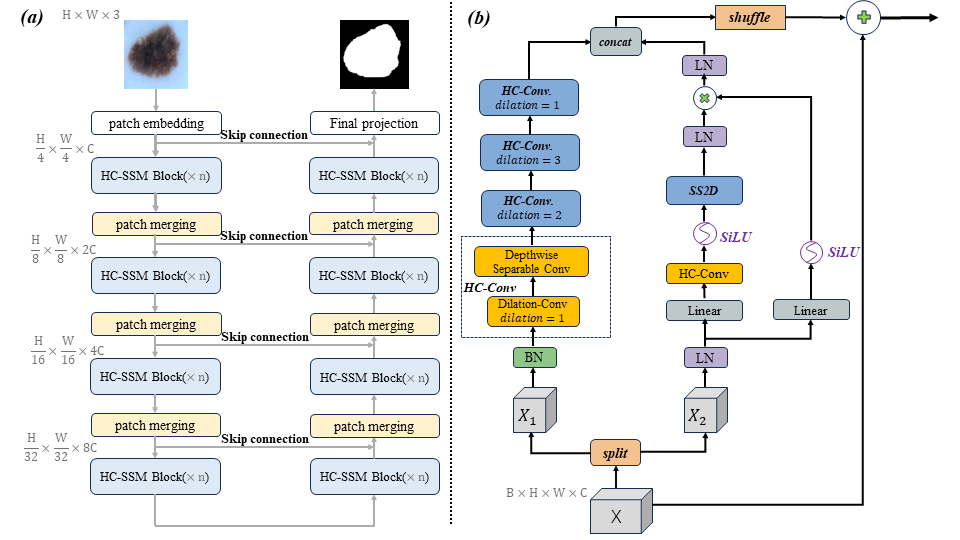}
    \caption{\textbf{(a) }Overall structure of HC-Mamba. \textbf{(b)} Overall structure of HC-SSM Bloc }
    \label{fig:2}
\end{figure}

The structure of HC-Mamba can be described as patch embedding layer, HC-SSM Block and patch merging layer. the model architecture is shown in Figure \ref{fig:2}(a).
In the HC-Mamba , the Patch Embedding layer first partitions the input image \(x \in \mathbb{R}^{H \times W \times 3}\) into non-overlapping blocks of size 4x4. This operation maps the dimensions of the image to \(C\) dimensions (typically \(C = 96\)), resulting in an embedded image \(x' \in \mathbb{R}^{\frac{H}{4} \times \frac{W}{4} \times C}\). Subsequently, \(x'\) undergoes a layer normalization to standardize the embedded image before entering the main backbone of the HC-Mamba. The backbone consists of four stages. In particular, after the output of the first three stages, a merging layer is used to reduce the height and width of the input features while increasing the number of channels. We employed [2, 4, 2, 2] HC-SSM blocks in the four stages, with each stage having [C, 2C, 4C, 8C] channels respectively.

\subsubsection{SS2D module}
SS2D module is the core of the HC-SSM block, which includes three key components: scan expansion, S6 block, and scan merging. Scan expansion decomposes the input image into independent sequences along four directions (up, down, left, and right), a step that ensures a wide spatial coverage of information and achieves multidirectional feature capture. Next, the S6 block uses a selectivity mechanism to impose choices on the parameters of the state-space model in order to accurately identify and extract the useful information while filtering out the irrelevant parts. Specifically, the block takes as input the feature format of $[B, L, D]$, where B is the batch size, L is the sequence length, and D is the feature dimension. The features are first transformed through a linear layer, after which the update and output equations in the state space model are applied to produce the final output features. Finally, a scan-and-merge operation reconfigures these transformed sequences to produce an output image that matches the dimensions of the original input image. Through this subtle series of operations, the SS2D module provides powerful feature extraction and processing capabilities for the HC-SSM block.
\subsubsection{HC-SSM Block}
\begin{figure}[H]
  \centering
  \includegraphics[scale=0.5]{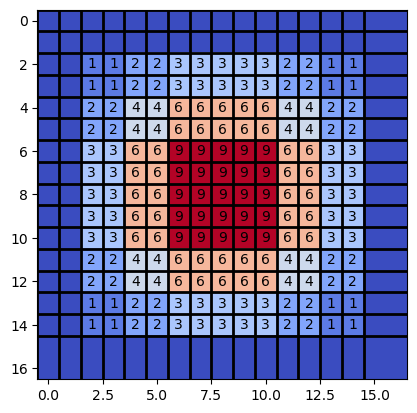}
  \hspace{1in}
  \includegraphics[scale=0.5]{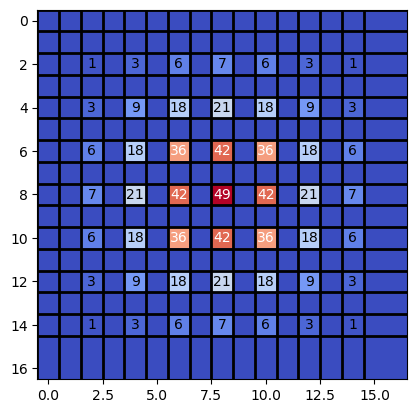}
  \caption{Comparison diagram between expansion rate of 1,2,3 \textit{(left)} and expansion rate of 2,2,2 \textit{(right)}}
  \label{fig:3}
\end{figure}

HC-SSM block is the core module of HC-Mamba, as shown in Figure \ref{fig:2}(b). We propose a two-branch feature extraction module based on SS2D. First, the module input is split into two sub-inputs of equal size using the channel split operation. Then, the two sub-inputs are fed into two branch modules, SSM branch and HC-Conv branch, respectively. In the SSM branch, the input undergoes a layer normalization and then enters the SS2D module, where the input features are first passed through a linear mapping for dimensionality enhancement, followed closely by a convolutional layer with depth-separable convolutions, which preserves the dimensionality and at the same time improves the localization processing of the features by grouping them. Then, the SiLU activation function is applied, a nonlinear transformation is introduced to enrich the model's expressiveness, and finally, the processed features are remapped to the original feature space to obtain the output of the SSM branch. In the HC-Conv branch, we introduce dilated convolution to expand the receptive field of the convolution kernel to capture a wider range of contextual information. This technique is particularly suitable for medical images, as it improves the model's ability to perceive structures at different scales of the image without losing image details. Meanwhile, we use an expansion strategy with an expansion rate of 1,2,3,1 to avoid the gridding effect that occurs with discontinuous data. Meanwhile, compared with the expansion rate of 2,2,2, the expansion rate of 1,2,3 strategy can ensure the continuity of the sensory field, an example is shown in Figure \ref{fig:3}.

\begin{figure}[H]
    \centering
    \includegraphics[width=0.5\linewidth]{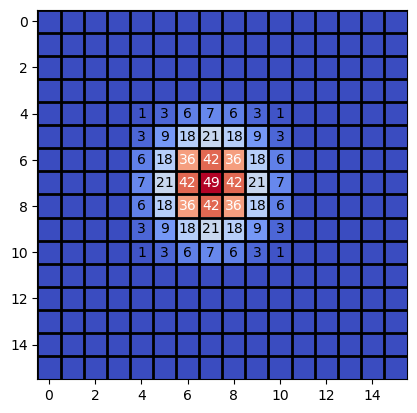}
    \caption{Receptive field diagram using three layers of ordinary convolution}
    \label{fig:4}
\end{figure}

In comparison to the use of three layers of normal convolution, a larger sensory field can be obtained, examples of which can be seen in Figure \ref{fig:4}.
Meanwhile, the use of a sawtooth-like expansion rate strategy(i.e., an expansion rate of 1,2,3,1) allows the refocusing of local features after multi-scale feature extraction and helps to maintain spatial continuity of features, while the use of a smaller expansion rate at the end of the sequence allows the model to refocus on smaller regions that may contain important information.

Finally, we merge the outputs of the two branches along the channel dimension of the feature map and use a parameter-free lightweight operation, the channel shuffle operation, to facilitate information interaction between the channels of the two sub-inputs.
\subsubsection{Loss function}
In the field of medical image segmentation, the most crucial evaluation metrics are the overlap between the segmentation results and the ground truth, the accuracy of the boundaries, and the similarity. Therefore, we designed a comprehensive weighted loss function that combines mIoU loss, Dice loss, and Boundary loss. The rationale for selecting these three loss functions is that each optimizes a distinct aspect of the segmentation task. The mIoU loss function evaluates the overlap of the segmented area, the Dice loss function measures the similarity of the segmented area, and the boundary loss function focuses on the accuracy of the segmentation boundaries. The combination of these three losses allows for a comprehensive optimization of the performance of the segmentation model.

Specifically, the loss function is expressed as follows:
\begin{equation}
\mathcal{L}=w_{\mathrm{mIoU}}\cdot\mathcal{L}_{\mathrm{mo}}+w_{\mathrm{Dice}}\cdot\mathcal{L}_{\mathrm{Dc}}+w_{\mathrm{Boundary}}\cdot\mathcal{L}_{\mathrm{Budr}}
\end{equation}

\begin{equation}
\left\{ 
    \begin{aligned}
        \mathcal{L}_{\mathrm{mo}} &= 1-\frac{1}{C}\sum_{i=1}^{C}\frac{\left|P_i\cap G_i\right|}{\left|P_i\cup G_i\right|} \\
        \mathcal{L}_{\mathrm{Dc}} &= 1-\frac{2\left|P\cap G\right|}{\left|P\right|+\left|G\right|} \\
        \mathcal{L}_{\mathrm{Budr}} &= \frac{1}{\left|B_P\right|}\sum_{p\in B_P}{\min_{q\in B_G}{d\left(p,q\right)}}+\frac{1}{\left|B_G\right|}\sum_{q\in B_G}{\min_{p\in B_P}{d\left(p,q\right)}}
    \end{aligned}
\right .
\end{equation}
where \( \mathcal{L}_{\mathrm{mo}} \), \( \mathcal{L}_{\mathrm{Dc}} \), and \( \mathcal{L}_{\mathrm{Budr}} \) represent the mIoU loss, Dice loss, and boundary loss, respectively. The terms \( w_{\mathrm{mIoU}} \), \( w_{\mathrm{Dice}} \), and \( w_{\mathrm{Boundary}} \) are the corresponding weight coefficients. The term \( B_P \) and \( B_G \) denote the boundary point sets of the predicted segmentation and the ground truth segmentation, respectively, and \( d(p, q) \) represents the Euclidean distance between point \( p \) and point \( q \).

The comprehensive loss function, which is the result of weighting and combining the three loss functions, is able to simultaneously optimize the overlap, similarity, and boundary accuracy of the segmentation regions,thereby improving the overall segmentation performance.

\section{Experiments}
\subsection{Datasets}
We conduct comprehensive experiments on HC-Mamba for medical image segmentation tasks. Specifically, we evaluate the performance of HC-Mamba on medical image segmentation tasks on the Synapse, ISIC17 and ISIC18 datasets.
\begin{itemize}
    \item \textbf{ISIC2017}:The ISIC2017 dataset contains three categories of diseases, melanoma, seborrheic keratosis, and benign nevus, 2,750 images, ground truth, and category labels. There are 2,000 images in the training set, 150 images in the validation set, and 600 images in the test set, and the color depth of the skin disease images is 24 bits, and the image sizes range from 767×576 to 6,621×4,441. The validation and test sets also include unlabeled hyperpixel images. The category labels are stored in tables and the datasets need to be preprocessed before training the model.
    \item \textbf{ISIC2018}:The ISIC2018 dataset contains different numbers of disease images for classification and segmentation, for the segmentation task, a total of 2,594 images were used as the training set, and 100 and 1,000 images were used as the validation and test sets, respectively. For the classification task, a total of 12,500 images were included, of which the training set contained a total of 10,015 images of 7 categories of diseases, namely actinic keratoses (327), basal cell carcinoma (514), benign keratoses (1,099), dermatofibromas (115), melanomas (1,113), melanocytic naevi (6,705), and vascular skin lesions (142). The seven classes of images in the classification task dataset are mixed in the same folder, and the labels are stored in tables that require preprocessing.
    \item \textbf{Synapse}:The Synapse Multi-Organ Segmentation dataset contains CT images from over 30 patients for abdominal multi-organ segmentation. This dataset includes segmentation labels for organs such as the liver, spleen, and pancreas, totaling over 4200 images. The images are of uniform resolution and provide accurate labeling of organ contours to support the needs of medical image processing and deep learning research. The dataset is widely used for training and evaluation of medical image segmentation algorithms, especially for improving the accuracy and efficiency of automatic segmentation.
\end{itemize}
\subsection{Results}

We compare HC-Mamba with some state-of-the-art models and some recent mamba-based model, presenting the experimental results in Table \ref{tab:isic} and \ref{tab:synapse}. 

For the ISIC2017 and ISIC2018 datasets, HC-Mamba performs well on mIoU and Dice compared to other models. Specifically, HC-Mamba has a 0.29\% and 0.2\% advantage over MedMamba on mIoU and Dice, respectively, while it has a 0.9\% and 1.48\% advantage over Unet on mIoU and Dice.
\begin{table}[!th]
	\setlength\tabcolsep{3pt}
	\renewcommand\arraystretch{1.25}
	\scriptsize
	\caption{Comparative experimental results on the ISIC17 and ISIC18 dataset. (\textbf{Bold} indicates the best.)}
	\begin{center}
		\begin{tabular}{c|c|ccccc}
			\hline
			\textbf{Dataset} &\textbf{Model}          & \textbf{mIoU(\%)$\uparrow$}  & \textbf{DSC(\%)$\uparrow$}   & \textbf{Acc(\%)$\uparrow$}   & \textbf{Spe(\%)$\uparrow$}   & \textbf{Sen(\%)$\uparrow$}   \\ \hline
			\multirow{5}{*}{ISIC17} &UNet\cite{ronneberger2015u}  &76.98 &85.99 &94.65 &97.43 & 86.82  \\
			&UTNetV2\cite{utnetv2}                & 76.35          & 86.23          & 94.84          & 98.05          & 84.85          \\
			&TransFuse\cite{zhang2021transfuse}               & 77.21          & 86.40          & 95.17          & 97.98          & 86.14 \\
            &MALUNet\cite{malunet}  &76.78 &87.13 &95.18 &\textbf{98.47} &84.78 \\
			&VM-UNet\cite{ruan2024vmunet} &77.59 &87.03 &\textbf{95.40} &97.47 &86.13   \\
                &MedMamba\cite{yue2024medmamba} &77.82&87.18&95.01 &97.50 &86.62 \\
                &HC-Mamba &\textbf{77.88}&\textbf{87.38}&95.17 &97.47 &\textbf{86.99} \\ \hline\hline
			\multirow{8}{*}{ISIC18}&UNet\cite{ronneberger2015u}                     & 77.86          & 87.55          & 94.05          & 96.69          & 85.86          \\
			&UNet++ \cite{unet++}                & 76.31          & 85.83          & 94.02          & 95.75          & 88.65          \\
			&Att-UNet \cite{attentionunet}             & 76.43          & 86.91          & 94.13          & 96.23          & 87.60          \\    
			&SANet \cite{sanet}                 & 77.52          & 86.59          & 94.39          & 95.97          & 89.46 \\
			&VM-UNet\cite{ruan2024vmunet} &77.95 &87.61  &94.13   &96.94&85.23 \\
            &MedMamba\cite{yue2024medmamba} &78.13&87.78&94.23 &95.68 &\textbf{89.74 }\\
            &\textbf{HC-Mamba} &\textbf{78.42}&\textbf{87.89}&\textbf{94.24}&\textbf{96.98}&88.90 \\ \hline
		\end{tabular}
		\label{tab:isic}
	\end{center}
\end{table}

For the Synapse datasets, HC-Mamba performs well on DSC and HD95 compared to other models. Specifically, HC-Mamba has a 0.31\% and 0.5\% advantage over MedMamba and VM-Unet on Dice.
\begin{table}[!th]
	\setlength\tabcolsep{1.2pt}
	\renewcommand\arraystretch{1.6}
	\tiny
	\caption{Comparative experimental results on the Synapse dataset. DSC of every single class is also reported. (\textbf{Bold} indicates the best.)}
	\begin{center}
		\begin{tabular}{c|cc|cccccccc}
			\hline
			\textbf{Model}          & \textbf{DSC$\uparrow$} & \textbf{HD95$\downarrow$} & \textbf{Aorta} & \textbf{Gallbladder} & \textbf{Kidney(L)} & \textbf{Kidney(R)} & \textbf{Liver} & \textbf{Pancreas} & \textbf{Spleen} & \textbf{Stomach} \\ \hline
			V-Net \cite{vnet}                   & 68.81            & -                 & 75.34          & 51.87                & 77.10              & 80.75     & 87.84          & 40.05             & 80.56           & 56.98            \\
   DARR \cite{DARR}  &69.77 &- &74.74 &54.47&72.31 &73.24 &94.08 &54.18 &89.20&45.96 \\
			UNet \cite{ronneberger2015u}                  & 76.85            & 42.70& 89.07 & \textbf{69.72}       & 77.77              & 68.60              & 93.43          & 53.98             & 86.67           & 75.58            \\
			Att-UNet \cite{attentionunet}                & 77.77            & 39.02& 89.55& 68.88                & 77.98              & 71.11              & 93.57          & 58.04             & 87.30           & 75.75            \\
			TransUnet \cite{chen2021transunet}              & 76.38& 34.69& 86.13& 62.03& 80.77& 75.92& 92.98& 54.76& 83.98& 74.52\\
Swin U-Net \cite{cao2022swin}  &77.15&29.55&83.49&64.55&81.30&77.63&92.31&54.60&88.68&74.62\\
   VM-UNet\cite{ruan2024vmunet} &79.08&32.21&84.40&68.41&83.16&\textbf{80.74}&92.07&56.90&87.51&79.42\\
   MedMamba\cite{yue2024medmamba} &79.27&28.15&86.23&67.53&81.22&78.49&94.89&\textbf{58.57}&89.29&78.12\\
   \hline
   Hc-Mamba &\textbf{79.58}&\textbf{26.34}&\textbf{89.93}&67.65&\textbf{84.57}&78.27&\textbf{95.38}&52.08&89.49&\textbf{79.84}\\
   \hline 
		\end{tabular}
		\label{tab:synapse}
	\end{center}
\end{table}

\subsection{Ablation experiments}
We compare HC-Mamba with and without Dilated convolution and depthwise separable convolution(DW convolution), presenting the experimental results in Table \ref{tab:Ablation}.Compared with model without Dilated convolution and depthwise separable convolution, HC-Mamba has only 12M parameters, a reduction of nearly 60\%, while maintaining the same high level of performance.

\begin{table}[!th]
    \centering
    \setlength\tabcolsep{3pt}
	\renewcommand\arraystretch{1.25}
 	\caption{Ablation studies on dilated convolution and depthwise separable convolutions.}
\begin{tabular}{c|cc|c}
\hline
\multirow{2}{*}{Convolution Method} & \multicolumn{2}{c|}{Evaluation} & \multicolumn{1}{c}{parameter count} \\ 
             & mIoU(\%)$\uparrow$         & DSC(\%)$\uparrow$     & Count(M)$\downarrow$              \\ \hline
 - &77.59&87.03&27.43  \\
Dilated convolution     & 78.40& 87.85& 24.68              \\ 
DW convolution     & 77.60& 87.02& 13.06              \\ 
Both &78.42& 87.89&13.88               \\  \hline
\end{tabular}
    \label{tab:Ablation}
\end{table}

\section{Discussion}
We propose HC-Mamba, a SSM model based on optimized convolution of multiple medical images. Its performance on medical image segmentation tasks is due to some of the current state-of-the-art models and some of the recent Mamba-based models. 

We introduce the technique of dilated convolution in the HC-Mamba model. Dilated convolution technique enables the model to capture a more extensive range of contextual information without increasing the computational cost by extending the perceptual field of the convolution kernel. This technique is particularly well-suited to medical images because it enhances the model's ability to perceive structures at different scales of the image without losing image details. Concurrently, by inputting the features generated by the dilated convolution into SSM, the state transition capability of SSM can be utilized to enhance the spatial correlation between the features, thus compensating for the discontinuities introduced due to the voids, which is one of the reasons for the excellent performance of HC-Mamba on medical images.

In addition, the HC-Mamba model employs depthwise separable convolutions, a convolution method that decomposes the traditional convolution operation into two parts: depthwise convolution and pointwise convolution, significantly reducing the number of parameters and the computational power of the model. By combining dilated convolution and depthwise separable convolutions, HC-Mamba is able to process large-scale medical image data at a much lower computational cost while maintaining a high level of performance. Compared with existing Mamba-based segmentation models, such as VM-Unet, which has nearly 30M parameters, and MedMamba, which has nearly 25M parameters, HC-Mamba has only 13M parameters, a reduction of nearly 50\%, while maintaining the same high level of performance, which provides a better basis for deploying it on lower-end devices.
\section{Conclusion}
We propose a Mamba for medical image segmentation (HC-Mamba) that incorporates multiple convolutional approaches optimized for medical images, and the HC-SSM module for enhancing the model's ability to extract features. The proposed method achieves excellent performance compared to some state-of-the-art models and some recent mamba-based model.  The results show that HC-Mamba has excellent performance in medical image segmentation tasks. In addition, compared to various architectures widely used in medical image segmentation tasks, HC-Mamba demonstrates strong competitiveness.
In addition, we summarize our future work into the following points: 1) We will further explore and test the potential of HC-Mamba on medical datasets obtained from other imaging technologies. 2) We will further use explainable artificial intelligence to analyze HC-Mamba’s decision-making mechanism. 

\bibliographystyle{unsrt}  
\bibliography{references}  


\end{document}